\documentclass[12pt,a4paper,final]{iopart}
\usepackage{iopams}  
\usepackage{graphicx}
\usepackage[breaklinks=true,colorlinks=true,linkcolor=blue,urlcolor=blue,citecolor=blue]{hyperref}

\newcommand{\erfc}{\textrm{erfc}}

\newcommand{\vc}{\mathbf}
\usepackage{graphicx}
\usepackage{color}

\begin{document}


\title[Extensions and Applications of the Bohm Criterion]{Extensions and Applications of the Bohm Criterion}


\author{Scott D.\ Baalrud, Brett Scheiner}
\address{Department of Physics and Astronomy, University of Iowa, Iowa City, Iowa 52242, USA}

\author{Benjamin Yee, Matthew Hopkins and Edward Barnat}
\address{Sandia National Laboratories, Albuquerque, New Mexico 87185, USA}

%
%



\begin{abstract}

The generalized Bohm criterion is revisited in the context of incorporating kinetic effects of the electron and ion distribution functions into the theory. The underlying assumptions and results of two different approaches are compared: The conventional `kinetic Bohm criterion' and a fluid-moment hierarchy approach. The former is based on the asymptotic limit of an infinitely thin sheath ($\lambda_D/l =0$), whereas the latter is based on a perturbative expansion of a sheath that is thin compared to the plasma ($\lambda_D/l \ll 1$). Here $\lambda_D$ is the Debye length, which characterizes the sheath length scale, and $l$ is a measure of the plasma or presheath length scale. The consequences of these assumptions are discussed in terms of how they restrict the class of distribution functions to which the resulting criteria can be applied. Two examples are considered to provide concrete comparisons between the two approaches. The first is a Tonks-Langmuir model including a warm ion source [Robertson 2009 \textit{Phys.\ Plasmas} {\bf 16} 103503]. This highlights a substantial difference between the conventional kinetic theory, which predicts slow ions dominate at the sheath edge, and the fluid moment approach, which predicts slow ions have little influence. The second example considers planar electrostatic probes biased near the plasma potential using model equations and particle-in-cell simulations.  This demonstrates a situation where electron kinetic effects alter the Bohm criterion, leading to a subsonic ion flow at the sheath edge. 


\end{abstract}



\vspace{2pc}
\submitto{\PPCF}


\maketitle



\section{Introduction}

The Bohm criterion is a statement that ions flow supersonically into ion sheaths: $V_i \geq c_s \equiv \sqrt{T_e/m_i}$~\cite{tonk:29,bohm:49}. This has become an important result because the marginal condition (sonic speed) provides a robust rule of thumb for most plasmas. It greatly simplifies modeling by providing a boundary condition without the need for a detailed description of the often complicated ion dynamics occurring throughout the plasma-boundary transition region (presheath). It is a ubiquitous result applied throughout plasma physics. A few examples are probe diagnostics~\cite{hers:89}, materials processing~\cite{lieb:05}, global plasma models~\cite{gudm:01}, fusion plasma-wall interactions~\cite{stan:84,cohe:04}, and space plasmas~\cite{robe:13}.  It is especially useful in weakly collisional situations where there is a large separation of scale between the sheath and plasma (or presheath). In this situation ion dynamics are typically well approximated as ballistic through the sheath, so the sheath edge properties of the ions can be advanced to the boundary without requiring a detailed model of the sheath. 

Although it gives accurate results in most circumstances, the Bohm criterion is not a physical law in the sense that exceptions exist and modifications are sometimes required. This was emphasized in Bohm's original paper~\cite{bohm:49}: ``This criterion is not exact, however, because it depends somewhat on the distribution of ionic velocities at the sheath edge. Yet, within 20 or 30 percent it applies in practically all cases.'' His point can be readily inferred from the fact that the expression concerns fluid variables, indicating an approximation of local thermodynamic equilibrium amongst individual species. However, many effects can modify the distribution functions, including absorption by boundaries, secondary electron emission and collisions, to name a few. It is important to understand how such kinetic effects may alter the Bohm criterion. 

Generalized Bohm criteria relate properties of the electron and ion distribution functions at the edge of a stable sheath. There are two key aspects to obtaining such a criterion: (1) identifying the sheath edge, and (2) applying a plasma model to relate electron and ion properties at this location. The concept of a sheath edge is based on Langmuir's two-scale description of neutral plasma and non-neutral sheath~\cite{lang:23}. Subsequent analyses have shown that a transition region can also be identified, which often depends on the ion collision mean free path~\cite{riem:91}. Features of this more detailed description have been measured experimentally~\cite{oksu:02}.  In the present work, we limit the discussion to weakly collisional situations where the standard two-scale picture is accurate. Our focus is on the second aspect. For this, Bohm applied a seemingly simple model taking the electron density to be Boltzmann distributed [$n_e = n_o \exp (-e \phi/T_e)$] and ions to be monoenergetic ($V_i = \sqrt{2e \phi/M_i}$) without sources or sinks ($n_iV_i =$ constant). Although this model is quite restrictive, the Bohm criterion is so useful because it remains accurate even when these underlying assumptions are inaccurate. This has motivated the search for a kinetic Bohm criterion that accounts for arbitrary distribution functions. Such a generalization is useful both from the standpoint of understanding the limitations of Bohm's criterion, and for providing a practical theory for cases where modifications are required.

An expression that is commonly called `the kinetic Bohm criterion' is \cite{lieb:05}
\begin{equation}
\frac{1}{M_i} \int d^3v \frac{f_i (\vc{v})}{v_z^2} \leq - \frac{1}{m_e} \int d^3v \frac{1}{v_z} \frac{\partial f_e (\vc{v})}{\partial v_z}  . \label{eq:kbc}
\end{equation}
This result stems from a progression of work starting from Harrison and Thompson's solution of the Tonk's-Langmuir problem~\cite{harr:59}, which was used to generalize the ion term in Bohm's result. More recent developments~\cite{riem:81} have also relaxed Bohm's assumption for the electron distribution, leading to Eq.~(\ref{eq:kbc}). Equation~(\ref{eq:kbc}) is used in many areas of plasma physics \cite{gudm:01,stan:84,cohe:04,emme:80,biss:87,sche:88,vand:91,biss:89,alle:09,ahed:10,rizo:14}. It is a successful generalization in the sense that: (a) it contains Bohm's result ($V_i \geq c_s$) if his assumptions of a monoenergetic ion distribution [$f_i = n_i \delta (\vc{v} - \vc{V}_i)]$ and thermal electron distribution [$f_e = n_o \exp(-v^2/v_{Te}^2)/(\pi^{3/2} v_{Te}^3)$] are applied, and (b) it holds for an additional class of possible distribution functions. Some examples have been provided that solve for the ion distribution function throughout the sheath and presheath from model equations~\cite{emme:80,biss:87,sche:88,vand:91,tskh:14}. However, whereas Bohm's criterion is resilient to situations where the actual distribution functions deviate somewhat from the theoretical assumptions, Eq.~(\ref{eq:kbc}) tends to be sensitive to the assumptions of the underlying plasma model. 

This sensitivity restricts the class of distributions to which Eq.~(\ref{eq:kbc}) applies~\cite{hall:62,fern:05,baal:11,baal:12}. For instance, the ion distribution must be empty at $v_z=0$, otherwise the left side diverges. A flowing Maxwellian is a common example that cannot be treated by this theory.  This particular restriction results from the application of the \textit{asymptotic limit} $\lambda_D/l = 0$, rather than by treating $\lambda_D/l \ll 1$ as a perturbative expansion. Here, $\lambda_D$ is the Debye length and $l$ is the plasma (or presheath) length scale. A consequence is an infinitely thin sheath that allows no ions to escape. However, real sheaths have a finite thickness, allowing some ions to cross from the sheath to the plasma (for example, by ion collisions in the sheath or from ions born from ionization with enough energy to escape). Although the number of ions leaving the sheath is typically very small compared to the number flowing in, Eq.~(\ref{eq:kbc}) diverges when any number escape. This renders Eq.~(\ref{eq:kbc}) experimentally untestable. These difficulties have been pointed out before~\cite{hall:62,fern:05,baal:11,baal:12}. The question we address in this paper is, can the Bohm criterion be generalized in a different way to account for a broader class of distribution functions that are relevant to experiments? 

It was recently proposed that a simple way to do this is through the normal fluid moment procedure~\cite{baal:11}. In this method, a solution of the charge density from the usual two-fluid hierarchy of equations is inserted into the condition used to define the sheath edge. The fluid variables are related to the distribution functions through the standard integral definitions. For many cases, this is simply a restatement of Bohm's criterion $V_i \geq \sqrt{T_e/M_i}$ in terms of the moment definitions:
\begin{equation}
\int d^3v\, v_z f_i \geq \biggl[ \frac{1}{3} \frac{m_e}{M_i} \biggl( \int d^3v f_e \biggr) \biggl(\int d^3v\, v_z^2 f_e \biggr) \biggr]^{1/2}  .\label{eq:fm}
\end{equation}
Other cases can be more complicated, especially if temperature or pressure gradients are strong. At first look, this may not seem to be a desirable approach because the fluid moment hierarchy requires a closure that is not in general known. However, the approach is saved if one recalls that the usefulness of a Bohm criterion stems from relating the parameters used to define model distribution functions  (the Bohm criterion does not help if a full solution of the distribution functions is known). For instance, if the model for $f_i$ depends on the parameter $A$ and the model for $f_e$ on the parameter B, then the Bohm criterion may relate $A$ and $B$ (as a simple example, $A$ could be the ion flow velocity and $B$ the electron temperature, leading to $V_i \geq \sqrt{T_e/M_i}$). In this way, the distribution function model provides the closure. The number of moment equations required to obtain a closure is simply determined from the number of parameters in the model. Usually this is quite small (2 or 3) and the approach provides simple analytic predictions.

In this paper, we revisit the theoretical foundations of the Bohm criterion and compare the fluid moment approach with the standard kinetic Bohm criterion for two common situations where kinetic effects are significant. In this discussion, we find that it is conceptually useful to separate the concept of the sheath edge, which is determined entirely by the charge density, from the plasma model used to formulate a generalized Bohm criterion. Section~\ref{sec:se} reviews a variety of common methods for locating the sheath edge based on the charge density. Here it is stressed that the sheath edge is a meaningful concept so long as $\lambda_D/l \ll 1$.  Section~\ref{sec:kbc} details the two different methods of getting from a sheath edge definition to a kinetic Bohm criterion. Here, the differences between treating $\lambda_D/l$ in the asymptotic limit $\lambda_D/l = 0$, versus as a perturbation $\lambda_D/l \ll 1$ are emphasized. Finally, we consider two common examples that illustrate kinetic modifications to Bohm's original criterion, as well as differences between the two kinetic approaches. The first of these considers a theoretical model accounting for an ion source in the plasma-boundary transition region (Sec.~\ref{sec:ionc}), and the second considers particle-in-cell Monte-Carlo collision (PIC-MCC) simulations of an electrostatic probe biased near the plasma potential (Sec.~\ref{sec:probe}).

\section{Locating the sheath edge\label{sec:se}} 

The concept of a sheath edge is based on Langmuir's two-scale description separating quasineutral ``plasma'' and non-neutral ``sheath''~\cite{lang:23}. Although the reality is more complex, including a transition region~\cite{fran:04}, the two-scale concept remains useful and accurate so long as $\lambda_D/l \ll 1$. The scale $\lambda_D/l$ determines the accuracy of the two-scale approximation. We restrict our discussion to situations where this is small. In many derivations of the Bohm criterion (including Bohm's own work \cite{bohm:49}), the sheath edge is identified after the plasma model has been specified. For the present discussion, it is conceptually clarifying to separate the identification of the sheath edge, which depends only on the charge density, with the particular plasma model, which determines the generalized Bohm criterion. 

\begin{figure}
\begin{center}
\includegraphics[width=7cm]{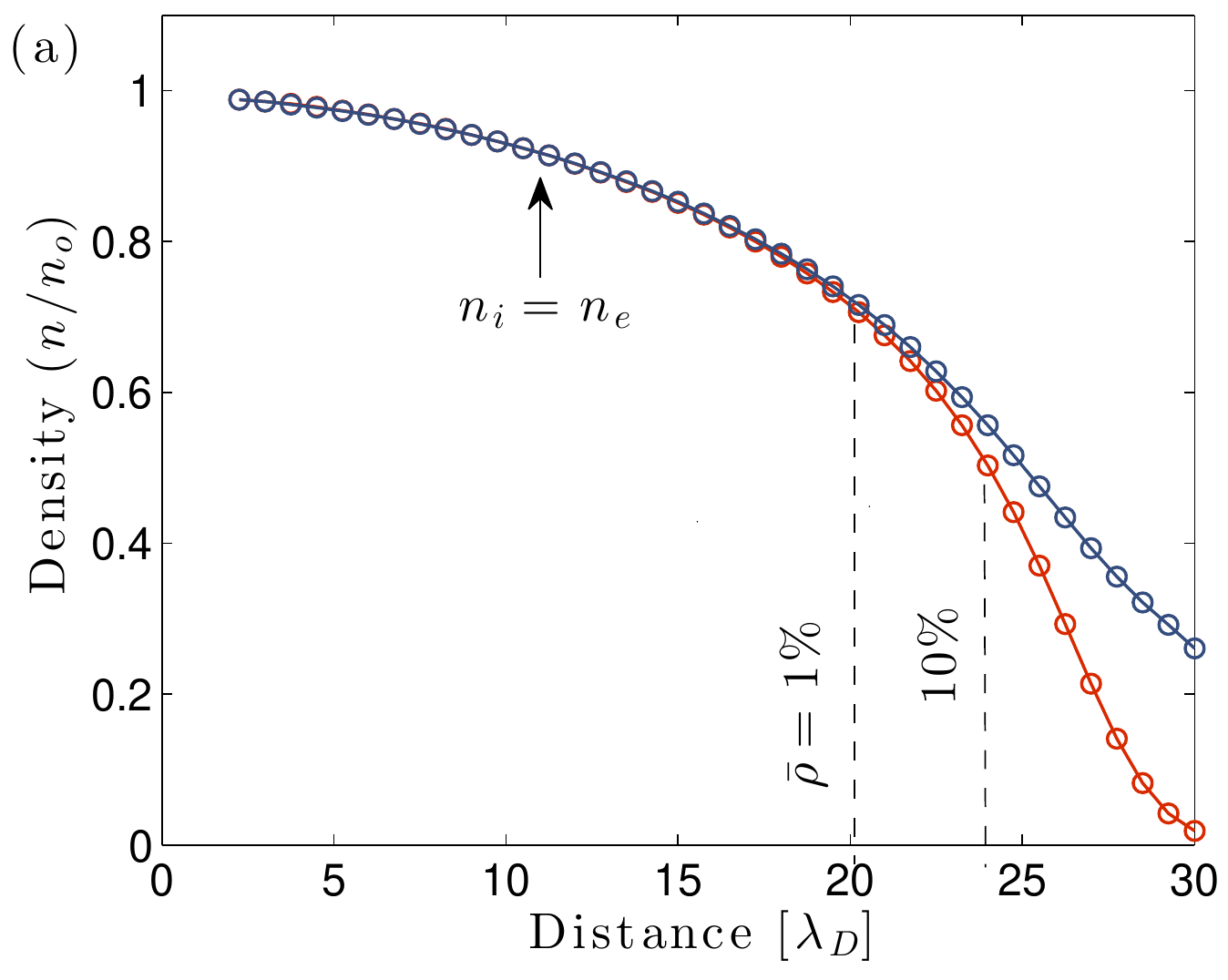}\\
\includegraphics[width=7cm]{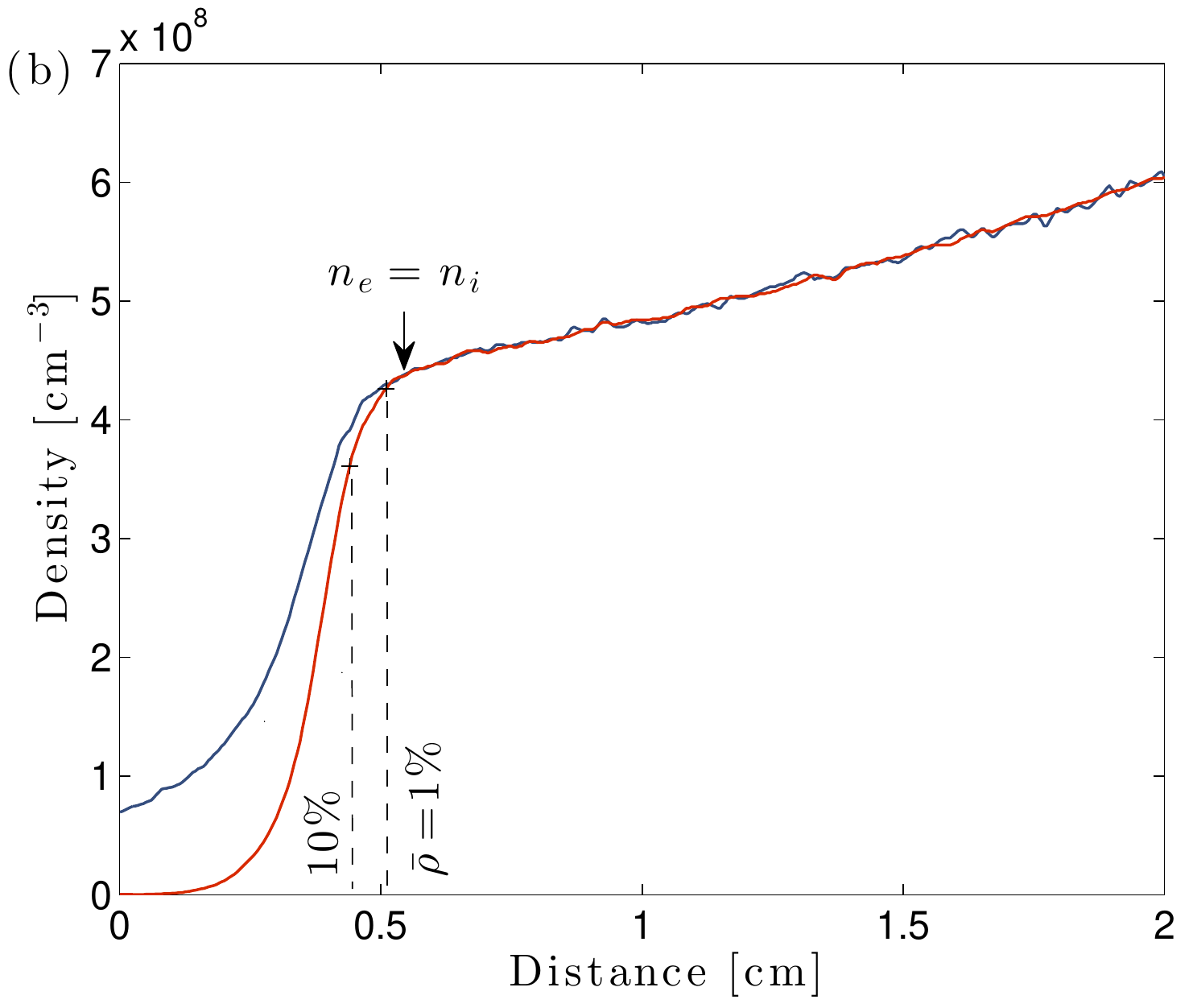}
\caption{Ion density (blue) and electron density (red) profiles from (a) Robertson's presheath model (see Sec.~\ref{sec:ionc}), (b) particle-in-cell simulations (see Sec.~\ref{sec:probe}). The location of normalized charge density is shown at 0, 1 and 10\% on each figure. }
\label{fg:density}
\end{center}
\end{figure}

The breakdown of quasineutrality in a sheath is abrupt if $\lambda_D/l \ll 1$. Two examples are shown in Fig.~\ref{fg:density}. This shows the electron and ion density profiles from the example problems discussed in Secs.~\ref{sec:ionc} and \ref{sec:probe}. Panel (a) shows an ion sheath calculated using Robertson's model~\cite{robe:09}, which is a Tonks-Langmuir model modified to include a warm Maxwellian ionization source. The free parameters are the domain length ($L$) in units of Debye length, and the ratio of electron temperature to ion source temperature $\tau = T_e/T_{is}$. Here $L=30$ and $\tau = 5$ were chosen. Panel (b) shows an example ion sheath from a PIC simulation that includes elastic ion collisions with background neutrals. In both examples, the location of the normalized charge density $\bar{\rho} = (n_i - n_e)/n_i$ is shown at three locations $\bar{\rho} = 0, 1$ and 10\%.

Since the presheath has a finite electric field, strict violation of charge neutrality is not necessarily associated with the sheath edge. For experimental or particle simulation data, such as panel b, the sheath edge can often be identified as the first position at which $n_e = n_i$ looking from the sheath toward the plasma. This is a viable approach when there is some low level of noise associated with the data, effectively identifying the sheath edge as the location where $\bar{\rho}$ is smaller than some fluctuation level associated with the data.  However, as panel (a) reveals, this location is both hard to identify and substantially further into the plasma than would usually be considered the sheath edge in model solutions with very low noise.  Instead, one can identify the sheath edge with the location where the charge density is smaller than some threshold. As an example, values of $\bar{\rho}$ are shown for 1\% and 10\%. Experimentally, a technique has been developed based on the inflection point of electron emitting probes operated in the limit of zero emission to identify where quasineutrality is violated~\cite{wang:06}. For theoretical development, further quantification is often desired. 

A common quantitative sheath criterion~\cite{riem:95} can be derived from a perturbative expansion of the charge density [$\rho = e (n_i - n_e)$] about the potential at the sheath edge ($\phi_o$): $\rho (\phi) = \rho (\phi_o) + d\rho/d\phi|_{\phi=\phi_o} (\phi - \phi_o) + \ldots$. The edge of an ion sheath is identified as the point where the charge density first becomes slightly positive: $\rho (\phi) - \rho (\phi_o) = \Delta \rho >0$. Taking this to be vanishingly small, the linear term in the expansion is of lowest order. Identifying the sheath edge with the marginal condition, the sheath criterion becomes $|d\rho/d\phi |_{\phi = \phi_o} \geq 0$. Substituting $d \rho/d\phi = - E^{-1} d\rho /dz$ (assuming a planar 1D model), this can be written in the familiar form 
\begin{equation}
\biggl| \frac{d n_i}{dz} \biggr| \leq \biggl| \frac{dn_e}{dz}  \biggr|. \label{eq:sc}
\end{equation} 
Equation~(\ref{eq:sc}) is commonly called the `sheath criterion' and it is often used to identify the sheath edge~\cite{hers:05}. It will be used in the next section when the different generalized Bohm criteria are obtained by inserting different plasma models for electron and ion density gradients. 

The salient feature here is that the sheath edge is a concept associated with the charge density, which does not require reference to the plasma model. This distinction is often not made explicit. Instead, it is common to identify the sheath edge after the plasma model has been chosen. For example, Bohm suggested using the location where $V_i = c_s$ to locate the sheath~\cite{bohm:49}, but this is obviously not adequate when seeking a generalized theory. Another way is to use the Child-Langmuir expression for the sheath thickness 
\begin{equation}
\frac{s}{\lambda_D} = \frac{\sqrt{2}}{3} \biggl( \frac{2 e \Delta \phi_s}{T_e} \biggr)^{3/4} ,
\end{equation}
and define the sheath edge as being a distance $s$ away from the boundary. Here, $\Delta \phi_s$ is the sheath potential drop. However, this also relies on a plasma model, including the concept of Bohm current for ions entering the sheath.\footnote{The ``Bohm'' criterion was actually appreciated much earlier by Langmuir. Perhaps it should be called the Langmuir criterion, but we apply the standard terminology.} An alternative method is to identify the sheath edge through a singularity in model plasma equations~\cite{riem:95}. However, the sheath edge has a physical significance regardless of the assumptions made to obtain model equations.  It is also a physically meaningful concept in real plasmas, where the mathematical formality $\lambda_D/l \rightarrow 0$ that is often applied to define its location~\cite{caru:62,riem:12}, is not met. This distinction will prove consequential when comparing the two different kinetic Bohm criteria.

\section{From the sheath edge to a Bohm criterion\label{sec:kbc}}

A generalized Bohm criterion results from obtaining expressions for the density gradients from a plasma model and applying them to the sheath criterion from Eq.~(\ref{eq:sc}). This section discusses three different plasma models, but it is instructive to first consider what properties a generalized Bohm criterion should possess.  We expect that it should: (1) Provide information beyond the sheath edge definition itself, and (2) depend only on the local properties of the distribution functions at the sheath edge. The first requirement is that information must be provided beyond defining the breakdown of quasineutrality. The second requirement ensures that the criterion provides useful information without requiring a detailed solution of the presheath region. In practice, this typically means that the criterion depends on the parameters ($A, B, C$, etc.), but not their spatial derivatives. Next, we consider three examples.

\subsection{Bohm's derivation} 

Bohm derived his criterion by first inserting his plasma model into Poisson's equation, integrating, then taking an expansion similar to that used to derive Eq.~(\ref{eq:sc}) in order to identify the sheath edge~\cite{bohm:49}. Alternatively, the same result can be obtained by reversing the order of these steps and inserting his plasma model directly into Eq.~(\ref{eq:sc}). His model assumed Boltzmann distributed electrons $n_e = n_o \exp(-e \phi/T_e)$, and monoenergetic ions with velocity $V_i = \sqrt{2e\phi/M_i}$ without sources or sinks ($n_iV_i=\textrm{constant}$). This implies $dn_e/dz = -n_eE/T_e$ and $dn_i/dz = -n_iE/(M_i V_i^2)$. Inserting these into Eq.~(\ref{eq:sc}) and taking $n_i = n_e$ gives Bohm's inequality $V_i \geq \sqrt{T_e/M_i}$. 

\subsection{The conventional ``kinetic Bohm criterion''}

A kinetic Bohm criterion obtains the density gradients required in the sheath criterion of Eq.~(\ref{eq:sc}) from the plasma kinetic equation. For the 1D steady-state system considered here, this is 
\begin{equation}
v_z \frac{\partial f_s}{\partial z} + \frac{q_s}{m_s} E \frac{\partial f_s}{\partial v_z} = S_s  \label{eq:ke}
\end{equation}
where the subscript $s$ represents species (electrons or ions here) and $S_s$ represents the total collision operator that includes all possible collisions, including ionization sources and sinks, other plasma-neutral collisions (such as charge exchange), as well as collisions amongst charged particles. Equation~(\ref{eq:kbc}) can be derived by: (i) Dividing Eq.~(\ref{eq:ke}) by $v_z$, (ii) integrating the result over velocity to get an expression for the derivative of $n_s \equiv \int d^3v f_s$. Carrying these steps out for both electrons and ions gives 
\begin{equation}
\frac{1}{M_i} \int d^3v \frac{1}{v_z} \frac{\partial f_i}{\partial v_z} \leq - \frac{1}{m_e} \int d^3v \frac{1}{v_z} \frac{\partial f_e}{\partial v_z} + \frac{1}{eE} \int d^3v \frac{S_i - S_e}{v_z} . \label{eq:kbcs1}
\end{equation}
Finally, Eq.~(\ref{eq:kbc}) results from, (iii) neglecting the collision term on the right side of Eq.~(\ref{eq:kbcs1}), and (iv) integrating the left side by parts neglecting the velocity-space surface term. 

The combination of steps (i) and (iii) in this derivation imply the asymptotic limit $\lambda_D/l = 0$ is assumed. This is because $v_z$ can take all values, including zero. Dividing by zero here only leads to meaningful results when $f_s(v_z=0)=0$, which requires that no particles escape the sheath. In principle, keeping the entire expression, Eq.~(\ref{eq:kbcs1}), would be mathematically valid, but the divergent source term could not be dropped in this case. A different viewpoint for justifying this approximation is that the electric field becomes singular at the sheath edge in the asymptotic limit $\lambda_D/l = 0$, justifying the neglect of the source term. Once again, this requires the mathematical idealization of an infinitely thin sheath.  Step (iv) of this derivation has also been shown by a few different authors to place further restrictions on the class of distribution functions that Eq.~(\ref{eq:kbc}) applies to~\cite{hall:62,fern:05,baal:11,baal:12}. This is because neglecting the velocity surface term in the integration-by-parts step requires $v_z^{-1}\partial f_i/\partial v_z$ to be continuously differentiable everywhere (including $v_z=0$). The implications of these aspects are discussed further in Sec.~\ref{sec:comp}.

\subsection{A criterion from the fluid-moment expansion} 

An alternative way to obtain $dn_s/dz$ is from the standard positive-exponent velocity moments (i.e., fluid moment expansion) of Eq.~(\ref{eq:ke})~\cite{baal:11}. This avoids placing restrictions on the distribution functions, but also requires that model distribution functions be specified before the hierarchy of fluid equations can be closed. The number of moment equations that are required is determined by the number of parameters in the model. 

The density moment of Eq.~(\ref{eq:ke}) gives
\begin{equation}
\frac{d}{dz} (n_s V_s) = G_s
\end{equation}
in which $G_s = \int d^3v S_s$ is a source or sink term (often from ionization). Density and flow velocity are now \textit{defined} by the distribution function using velocity-moment definitions: $n_s \equiv \int d^3v\, f_s$ and $V_s \equiv \int d^3v\, v_zf_s/n_s$. Inserting this into the sheath criterion from Eq.~(\ref{eq:sc}) gives
\begin{equation}
\frac{n_i}{V_i} \frac{dV_i}{dz} \leq \frac{n_e}{V_e} \frac{dV_e}{dz} + \frac{G_i}{V_i} - \frac{G_e}{V_e} . \label{eq:2p}
\end{equation}
The presence of spatial gradients in Eq.~(\ref{eq:2p}) signifies that this condition may not meet the ``local properties'' requirement for a Bohm criterion. However, for certain model distribution functions it is possible that the gradient terms may cancel after the models are inserted. A more broadly applicable result is obtained by taking one more moment.

The momentum moment of Eq.~(\ref{eq:ke}) is
\begin{equation}
m_s n_s V_s \frac{dV_s}{dz} = n_s q_s E - \frac{d p_s}{dz} - \frac{d \Pi_{zz,s}}{dz} + R_{z,s} \label{eq:mom}
\end{equation}
where $R_{z,s} = \int d^3v\, m_sv_zS_s$ is the friction force density resulting from collisions, and the scalar pressure and stress tensor component are defined by the usual moments of $f_s$: $p_s \equiv \int d^3v m_s v_r^2 f_s/3 \equiv n_s T_s$ and $\Pi_{s} \equiv \int d^3v m_s (\vc{v}_r \vc{v}_r - v_r^2 \mathcal{I} /3) f_s$, where $\vc{v}_r \equiv \vc{v} - \vc{V}_s$. Solving Eq.~(\ref{eq:mom}) for $dV_s/dz$ and putting the result in Eq.~(\ref{eq:2p}) gives
\begin{equation}
\sum_{s=e,i} \biggl[ \frac{ q_s^2 n_s - q_s ( n_s dT_s/dz + d \Pi_{zz,s}/dz + T_sG_s/V_s - R_s)/E}{m_s V_s^2 - T_s} \leq \frac{G_s}{EV_s} \biggr] . \label{eq:genbohm}
\end{equation}
Again, spatial derivates show up, which can be replaced by higher-order moment equations. However, a closure is ultimately required in order to remove these gradient terms. 

Although a completely general criterion cannot be written down in a closed form with this method, Eq.~(\ref{eq:genbohm}) can be simplified for most cases where ion sheaths are being considered. Under normal conditions, the scale length of temperature gradients, stress gradients and collision terms is much smaller than the gradient scale of the electrostatic potential at the sheath edge. In this case, all of the terms in Eq.~(\ref{eq:genbohm}) that are divided by $E$ can be neglected. Equation~(\ref{eq:genbohm}) then reduces to
\begin{equation}
V_i \geq \sqrt{\frac{T_e + T_i - m_e V_e^2}{M_i}}  \label{eq:vigb}
\end{equation}
where the kinetic form can be written simply by replacing the fluid variables by the integral definitions in terms of $f_i$ and $f_e$. Here, it was assumed that $m_eV_e^2 < T_e$.  Note that this approximation becomes rigorous in the asymptotic limit $\lambda_D/l =0$, where the sheath edge electric field becomes singular, but it also remains an accurate approximation when $\lambda_D/l$ is small and finite. Although Eq.~(\ref{eq:vigb}) applies to common ion sheath scenarios, we stress that a general criterion relies on the entire hierarchy (until a model is specified). We will consider an example where the predictions of Eqs.~(\ref{eq:genbohm}) and (\ref{eq:vigb}) differ in Sec.~\ref{sec:probe}.

\subsection{Comparing these approaches~\label{sec:comp}} 

Each of these approaches has advantages and disadvantages. An advantage of the conventional approach is that the result can be written in a simple compact form. A disadvantage is that it applies only to a restricted class of distributions that is purely theoretical, rendering it experimentally untestable. This is because real sheaths have a finite thickness, allowing for a finite population of ions to escape. The main advantage of the fluid moment approach is that it is quite general since it can accommodate any plasma (distribution function) model in principle.  An approximation arises in the identification of the sheath edge, for which the accuracy depends on $\lambda_D/l \ll 1$, but this is common to both approaches. A disadvantage is that the general expression depends on a hierarchy of equations. A closure can be obtained only after a plasma model has been specified. Although the vast majority of cases can be treated by the simple expression Eq.~(\ref{eq:vigb}), which neglects the terms divided by $E$ in Eq.~(\ref{eq:genbohm}), exceptions do exist (see Sec.~\ref{sec:probe}).

In practice, one wants to know which expression to apply to a given analytic model or simulation, or if one can constrain parameters of one species in an experiment based on a measurement of the other species. On the side of theoretical modeling, the conventional theory has been shown to hold for a few different models, including the Tonks-Langmuir problem where ions are born with zero energy~\cite{caru:62}, as well as a couple of other model source functions~\cite{emme:80,biss:87,sche:88,vand:91,tskh:14}. The Vlasov formulation is strictly valid only when electrons and ions are in equilibrium, otherwise,  there is an electron-ion collision term in Eq.~(\ref{eq:kbcs1}), but this not interesting from the standpoint of sheaths since ions flow relative to electrons. Since the conventional theory \textit{must} be applied outside of the regime where the underlying assumptions are strictly valid, it is important to address the question of the error introduced when doing this. Quantifying this error is often complicated by the dependence on the $v_{z}^{-1}$ moment of the total collision operator, which is also proportional to $f_s$ and/or $\partial f_s/\partial \vc{v}$. 

One surprising prediction of the conventional kinetic Bohm criterion is a high sensitivity to slow ions in the distribution function, suggesting that a small minority population can control the dynamics of the entire distribution. In contrast, the fluid moment approach is insensitive to a small population of slow ions. The fluid moment approach also allows one to cast the result in terms of the ion flow speed (defined through the moment definition), which provides a solid conceptual connection with Bohm's original formulation.

The role of slow ions is a definite prediction that highlights a stark contrast between the two approaches. As a simple example, consider a plasma model where electrons are Maxwellian and ions are a flowing Maxwellian truncated at a small velocity
\begin{equation}
f_i = \frac{n_i}{\pi^{3/2} v_{Ti}^3} \exp \biggl[ - \frac{(\vc{v} - V_{i} \hat{z})^2}{v_{Ti}^2} \biggr] H (v_z/v_{Ti} - \epsilon)  \label{eq:iontrunc}
\end{equation}
where $\epsilon > 0$, $H$ is the Heaviside step function, and $v_{Ti} = \sqrt{2T_i/m_i}$. The truncation avoids any divergences in the conventional theory, and considering small values of $\epsilon$ allows one to explore the role of slow ions in each approach. Using this in the fluid moment approach, in the form of Eq.~(\ref{eq:fm}), gives 
\begin{equation}
V_i \frac{1}{2} \biggl[ \erfc (\epsilon - u) + \frac{1}{\sqrt{\pi} u} e^{-(\epsilon - u)^2} \biggr] \geq \sqrt{\frac{T_e}{M_i}}
\end{equation}
where $u \equiv V_i/v_{Ti}$. For $\epsilon \ll 1$, this returns $V_i \geq \sqrt{T_e/M_i} +\mathcal{O}(\epsilon^2)$. Thus, a small population of slow ions is not significant in the fluid moment theory. In contrast, putting this plasma model into the conventional kinetic Bohm criterion, Eq.~(\ref{eq:kbc}), gives
\begin{equation}
\int_\epsilon^\infty dx \frac{e^{-(x-u)^2}}{x^2} \leq 2\sqrt{\pi} \frac{T_i}{T_e} .
\end{equation}
For $\epsilon \ll 1$, the integral on the left diverges at a rate $\propto 1/\epsilon$ (see figure 2 of \cite{baal:12}). Thus, the conventional theory predicts that the Bohm criterion is dominated by the behavior of slow ions~\cite{riem:12}. A drastically different physical picture is predicted by each approach. This was a contrived theoretical model, but the next two sections consider examples directly relevant to experiments.

\section{Role of ion generation in and near the sheath~\label{sec:ionc}} 

One important extension that kinetic theory provides is to go beyond Bohm's assumption of monoenergetic ions to consider the full ion distribution. Ions are born at different locations throughout the presheath in partially ionized plasmas. This affects the distribution by generating a tail that extends from the flowing bulk of the distribution down to low energy. The original Tonks-Langmuir solution modeled this by assuming ions are born with zero energy~\cite{tonk:29}. This is a reasonable approximation for modeling general features of the IVDF because the ion thermal speed is usually much smaller than the sound speed in low-temperature plasmas, i.e., $T_e\gg T_i$. For a more rigorous description, one can account for the fact that ions are born with a finite kinetic energy described by a distribution similar to that of the neutral gas. Several models have been developed to extend the Tonks-Langmuir model to include sources with an energy distribution~\cite{emme:80,biss:87,sche:88,vand:91,tskh:14,sher:01,kos:09,kos:13}

\begin{figure}
\begin{center}
\includegraphics[width=8.2cm]{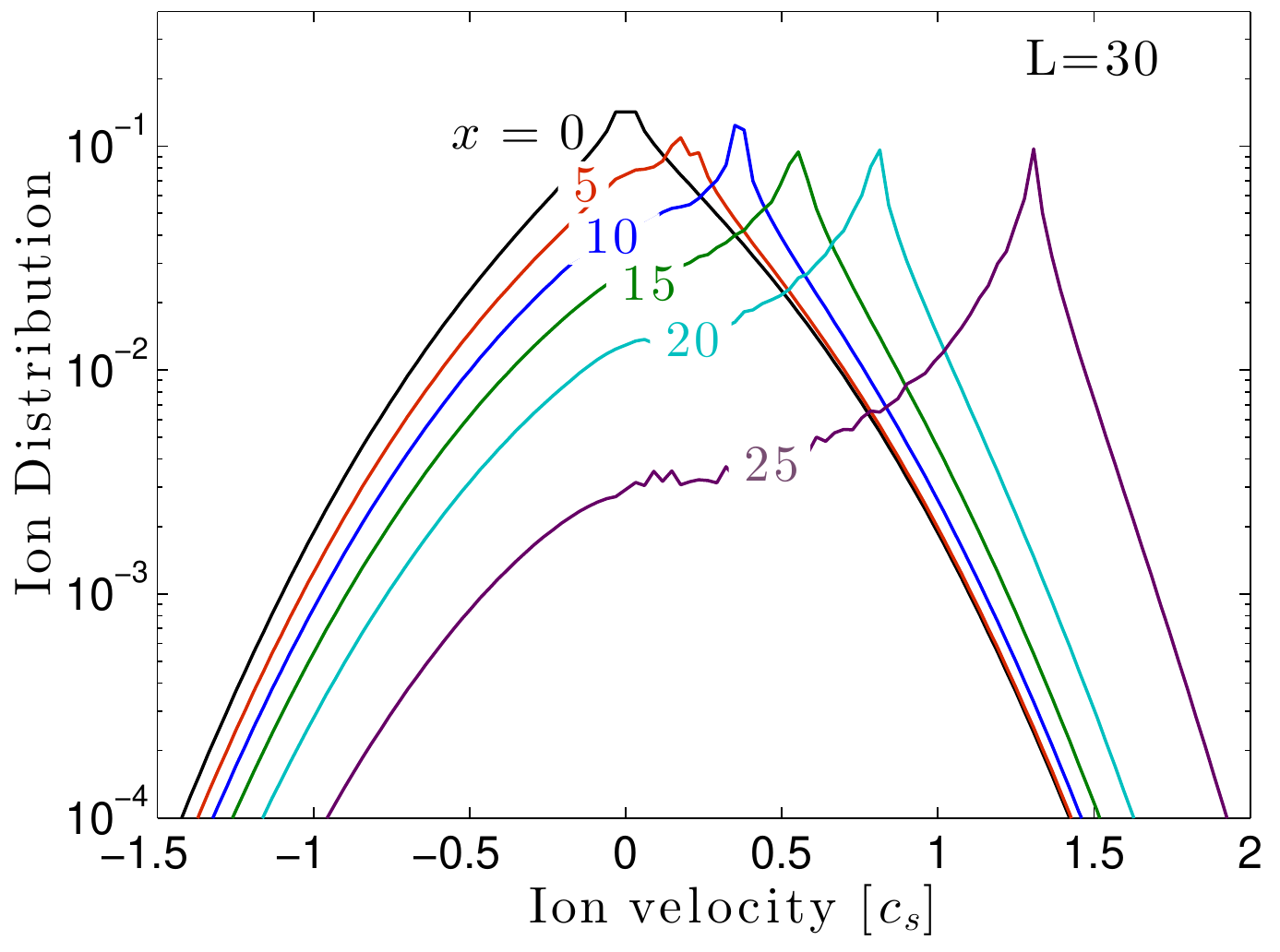}
\caption{Ion velocity distribution function from Robertson's sheath model from \cite{robe:09} shown at six locations throughout the domain. The sheath edge location is approximately $x=24$, as identified by $\bar{\rho}=0.1$ in Fig.~\ref{fg:density}. The chosen model parameters are the same as Fig.~\ref{fg:density}: $\tau = 5$ and $L=30$. The salient feature of this data is that a small population of ions can exit the sheath edge, as demonstrated by the finite values of the IVDF for negative velocities. The noise arises from finite differences taken in the calculation of $f_i$.}
\label{fg:robertson_ivdf}
\end{center}
\end{figure}

Here, we consider the plasma-sheath model developed by Robertson~\cite{robe:09} that models ion generation using a Maxwellian source. This model solves for the ion distribution function and electrostatic potential in a 1D domain assuming electrons are Boltzmann: $n_e = n_o \exp(-e\phi/T_e)$. The inputs to the theory are the domain length in units of Debye length $(L)$ and the ratio of electron temperature to ion source temperature ($\tau = T_e/T_{is})$. Details of the model are described in~\cite{robe:09}. 

\begin{figure}
\begin{center}
\includegraphics[width=8.2cm]{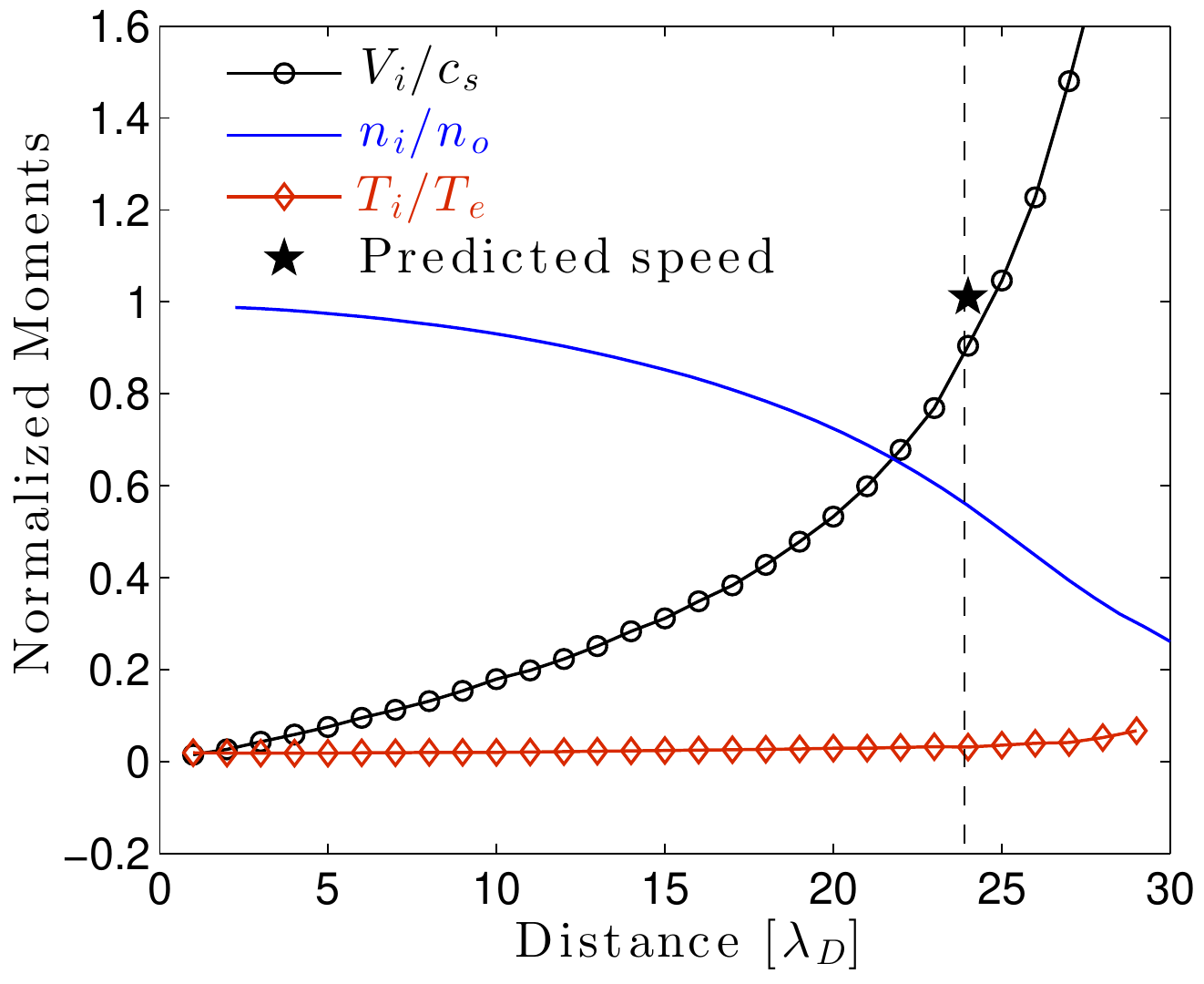}
\caption{Flow speed and temperature profiles calculated from the IVDFs shown in Fig.~\ref{fg:robertson_ivdf}, and the density profile from Fig.~\ref{fg:density}. The star shows the predicted ion speed at the sheath edge using the fluid moment approach.}
\label{fg:robertson_profile}
\end{center}
\end{figure}

Figure~\ref{fg:density} shows a solution of the density profile in the domain for parameters $L=30$ and $\tau=5$, and Fig.~\ref{fg:robertson_ivdf} shows the corresponding IVDFs at six locations spanning the presheath and sheath. As discussed in Sec.~\ref{sec:se}, the sheath edge is at approximately $x/L = 24$, within $\mathcal{O}(\lambda_D/L)$ accuracy, based on the location $\bar{\rho}=0.1$. Thus, the IVDF contains particles with $v_z \leq 0$ in both the presheath and sheath. Because $f_i(v_z=0)\neq 0$ at the sheath edge, the left side of Eq.~(\ref{eq:kbc}) diverges for this model, rendering the conventional kinetic criterion untenable. However, the fluid moment approach gives predictions essentially unchanged from Bohm's original criterion if the fluid variables are computed from the moments of the distribution functions, Eq.~(\ref{eq:fm}). The profile of the moment parameters are shown in Fig.~\ref{fg:robertson_profile}. Since temperature and stress gradients, as well as collision terms, are much smaller than the electric field terms at the sheath edge, Eq.~(\ref{eq:genbohm}) is well approximated by the simplified kinetic criterion in Eq.~(\ref{eq:vigb}). For a Maxwellian $f_e$, this gives the prediction $V_i \geq \sqrt{(T_e+T_i)/M_i} = c_s \sqrt{1+T_i/T_e} \simeq 1.01c_s$ ($T_i/T_e \simeq 0.02$ here). This prediction is only 1\% different than Bohm's (if the variables are treated through the moment definitions). Figure~\ref{fg:robertson_profile} shows that agrees well with the data within the uncertainty of locating the sheath edge, which is a few length units (Debye lengths) on the scale in this figure. 

This example demonstrates a situation where Bohm's original criterion is quite accurate despite the fact that the ion distribution deviates from his assumption of a monoenergetic beam. Slow ions play no special role in this example, in contrast to the prediction of the conventional kinetic Bohm criterion. Here, we considered one model for the IVDF, but the feature that some ions leave a sheath is common to all real experiments, and most particle simulations. A compilation of examples has been provided in Fig.~1 of \cite{baal:12}. Examples can be found from experimental measurements~\cite{bach:95,sade:97,oksu:01,seve:03,lee:06,clai:06,lunt:08,jaco:10,clai:12}, PIC simulations~\cite{milo:11} and hybrid simulations~\cite{sher:01,tsus:98}, all of which show IVDFs with a finite component of ions leaving the sheath. This is also a persistent feature of the IVDF near double layers~\cite{sun:05,scie:10,meig:05,baal:11b}, which is another sheath-like feature where the Bohm criterion applies.

\section{Probes biased near the plasma potential~\label{sec:probe}}

The previous section showed a situation treating modifications to Bohm's criterion from ion kinetic effects. In this section, we consider a common situation where electron kinetic effects arise: electrostatic probes biased near the plasma potential. As the sheath potential drop becomes smaller than the electron temperature, the electron distribution function is modified by wall losses, which can modify the Bohm criterion~\cite{baal:11,andr:70,loiz:11,jeli:11,camp:12,gyer:14}.

Electrostatic Langmuir probes are a staple of plasma diagnostics. This technique infers the electron temperature and density through the current collected as the probe bias is swept from negative to positive values (with respect to the plasma potential). In the ion saturation region (negative biases), it is typically assumed that the ions reach the probe with the Bohm current density, which is a direct consequence of the Bohm criterion. This is accurate if the probe bias is at least an electron temperature lower than the plasma potential, but electron kinetic effects become important for smaller biases because the wall provides a sink for much of the electron distribution. In this section, we consider the problem of small probe biases by comparing the kinetic Bohm criterion predicted from a model EVDF with PIC-MCC simulations. 

\begin{figure}
\begin{center}
\includegraphics[width=8.2cm]{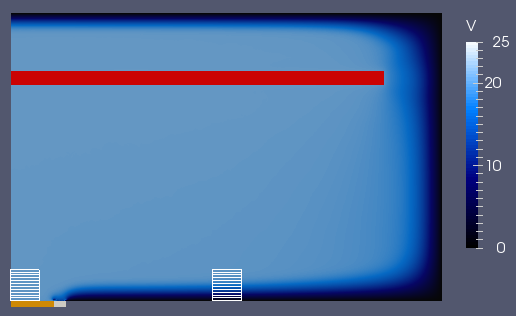}
\caption{Plasma potential contours throughout the domain of the 2D PIC-MCC simulations (shown by shading from light to dark). The biased electrode boundary is shown in the lower left corner. This is separated from the grounded wall boundary by a thin dielectric (shown in white). Plasma is generated in the thin box (shown in red). EVDFs were computed by averaging over the boxes shown in white. }
\label{fg:pic}
\end{center}
\end{figure}

The PIC-MCC simulations were carried out in the 2D domain shown in Fig.~\ref{fg:pic} using the code \textit{Aleph} for a helium plasma. \textit{Aleph} is an electrostatic plasma modeling code implementing Particle-In-Cell (PIC) and Direct Simulation Monte Carlo (DSMC) kinetic techniques. It utilizes unstructured meshes and can simulate 1D, 2D, and 3D Cartesian domains. The basic algorithm represents a plasma through evolving electrostatically coupled computational particles, each representing some number of real ones (e.g., e$^-$, He, or He$^+$). Because it tracks individual particles undergoing specific kinetic events (e.g., ionization), it is an excellent tool for simulating low temperature non-Maxwellian collisional plasmas.

A reflecting boundary condition was applied at the left domain wall. All others were perfectly absorbing. The electrode was biased +20V with reference to the grounded wall boundary. These two domains were separated by a thin dielectric, which was electrically floating, and was included to reduce abrupt discontinuities in the boundary conditions between the electrode and wall. Because the effective electrode surface area was sufficiently large, the plasma potential remained above the electrode potential to preserve global current balance~\cite{baal:07} ($A_E/A_w = 0.033$ in this case). The bulk plasma potential was found to be approximately 0.5V above the electrode potential (i.e., 20.5V).  Potential contours are shown throughout the domain in Fig.~\ref{fg:pic}. Plasma was generated at a constant production frequency (2.35$\times 10^9$ cm$^{-3}$s$^{-1}$) in the thin box shown in red. Both electrons and ions were generated with Maxwellian distributions at temperatures of 4 eV and 0.09~eV (1000 K) respectively. Electrons were observed to cool to approximately 1-2 eV in the bulk plasma, presumably due to loss of high energy electrons to the walls. The local plasma potential was enhanced by approximately 0.5V in the generation region, causing a slight drift of ions out of this region. However, this drift was substantially subsonic ($\lesssim 0.25 c_s$), so a presheath formed near both the electrode and wall sheaths. 

\begin{figure}
\begin{center}
\includegraphics[width=10cm]{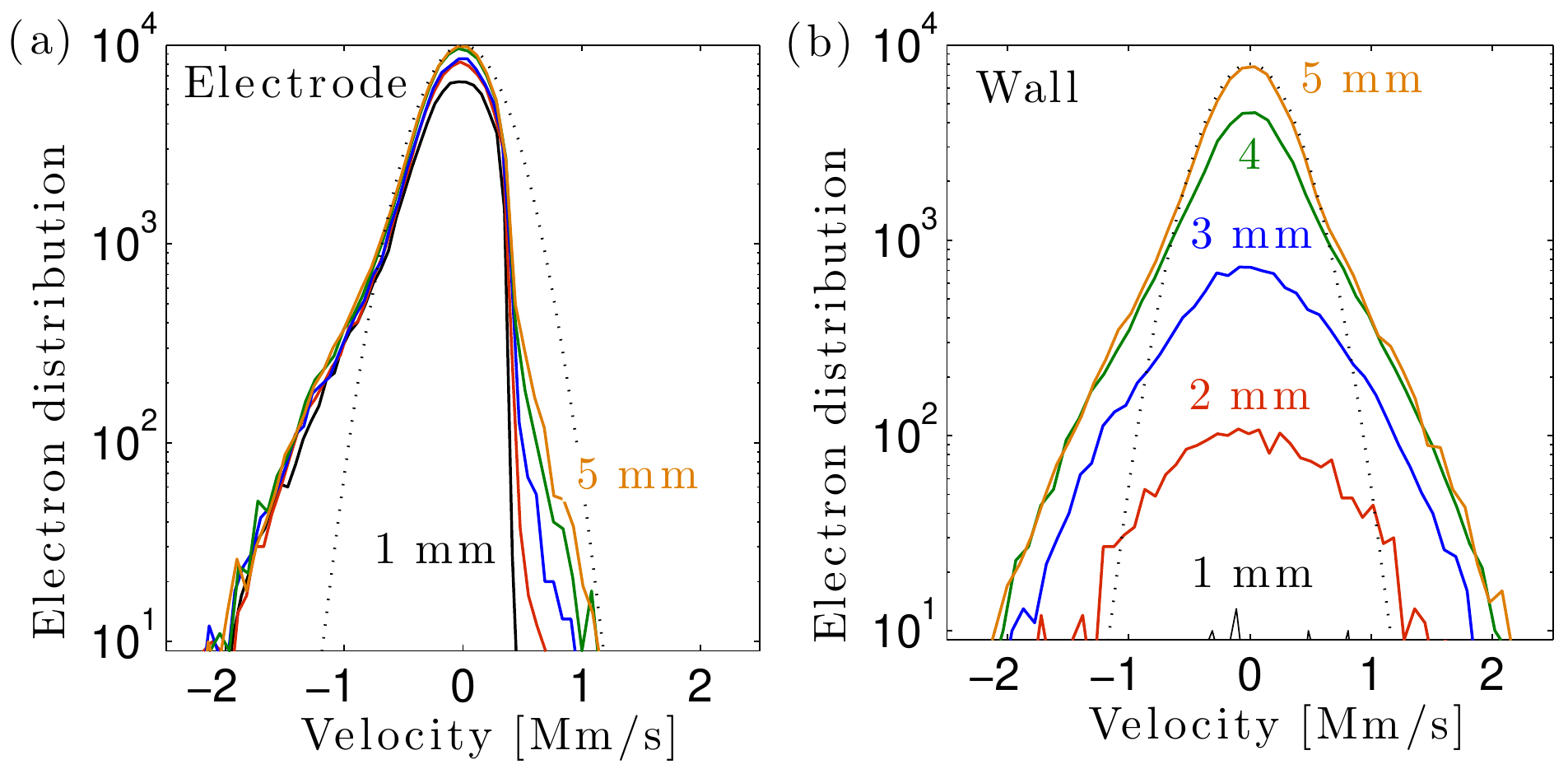}
\caption{Electron velocity distribution function in the two domains shown in Fig.~\ref{fg:pic}: (a) near the positive electrode, and (b) near the wall. Data are shown at 1, 2, 3, 4 and 5 mm from the boundary. The dotted line shows a Gaussian curve for reference. }
\label{fg:pic_evdf}
\end{center}
\end{figure}

\begin{figure}
\begin{center}
\includegraphics[width=8cm]{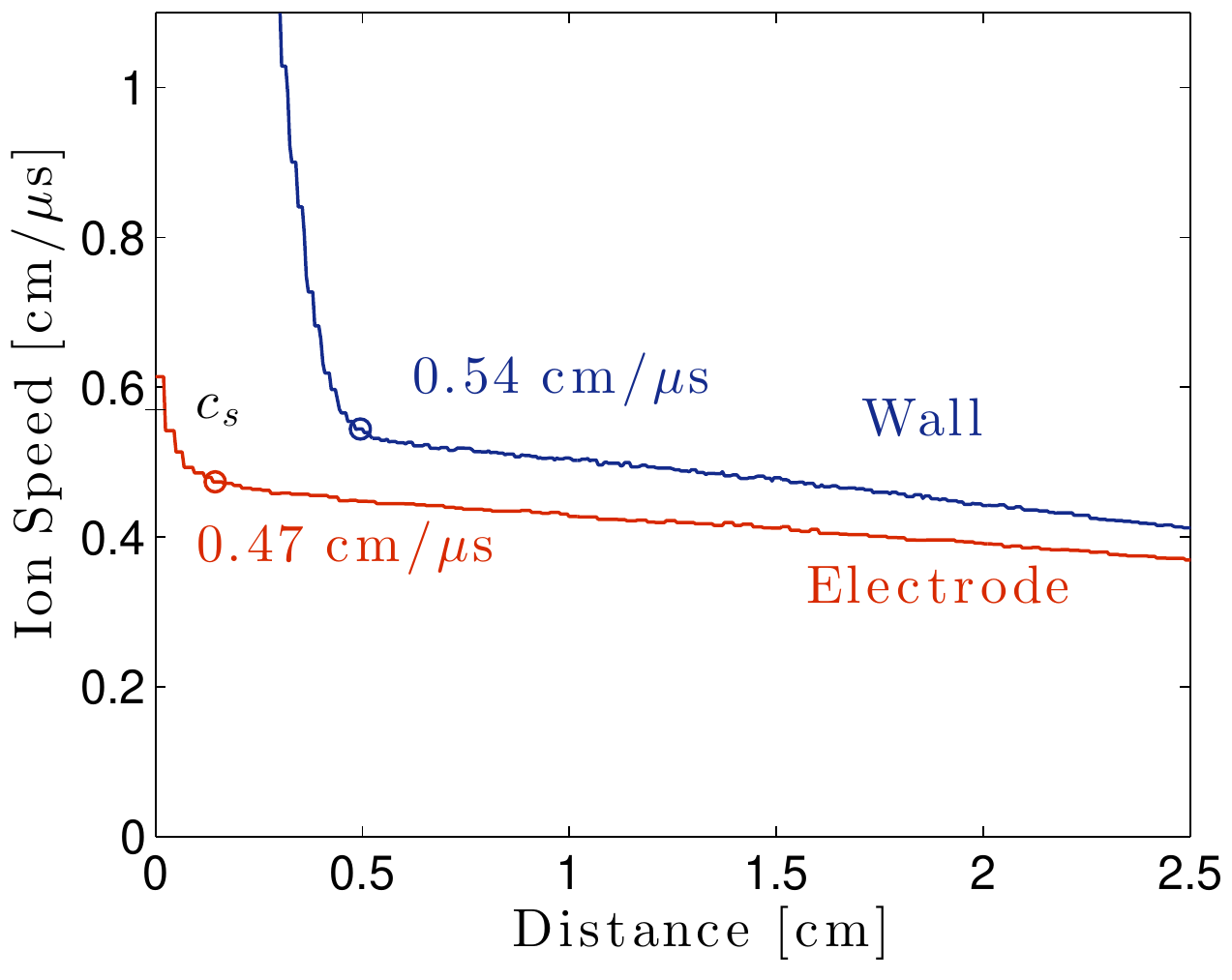}
\caption{Ion flow speeds in 1D domains centered at the electrode sheath (red) and wall sheath (blue). The ion flow speed at the sheath edge is indicated for each case, and is lower at the electrode sheath edge than the wall sheath edge.}
\label{fg:pic_vi}
\end{center}
\end{figure}

Here, we focus on comparing the Bohm criterion near the electrode sheath with the wall sheath. For this purpose, the electron distribution function has been computed in each region; see Fig.~\ref{fg:pic_evdf}. This was obtained by binning particle velocities from several different timesteps in the 1mm$\times$5mm boxes shown in Fig.~\ref{fg:pic} in both the electrode region and a region of the wall sheath in the center of the lower boundary. Figure~\ref{fg:pic_evdf} shows the data for the boxes centered at 1, 2, 3, 4 and 5 mm from each boundary. The salient feature of the EVDFs is that the tail is substantially depleted due to losses to the boundary in the vicinity of the electrode, but not the wall. This is a substantial departure from the assumed thermal electron distribution in Bohm's original formulation, motivating a kinetic analysis. 

Figure~\ref{fg:pic_vi} shows the ion flow speed profile along vertical segments from the bottom surface up a small distance from the domain, as shown in Fig.~\ref{fg:pic}. The electrode sheath data is taken along the left hand boundary, representing the center of the electrode, and the wall sheath data from the center of the domain shown, representing a characteristic ion sheath (corresponding to the regions used in Fig.~\ref{fg:pic_evdf}). This shows a slower ion speed at the edge of the electrode sheath than the wall sheath. The sound speed is estimated to be approximately 0.57 cm/$\mu$s based on the bulk electron temperature of approximately 1.4 eV. This temperature was computed from the moment definition for the 2D domain: $T_e = \int d^3v m_e (v_{r,x}^2 + v_{r,y}^2)f_e/2$.  The sheath edge positions were determined by the method of finding the first location where $n_e = n_i$ looking from the sheath to the plasma (as explained in Sec.~\ref{sec:se}).  The ion flow speed at the wall sheath edge was 0.54 cm/$\mu$s, which is consistent with the usual Bohm criterion within the error of identifying the sheath edge. The ion flow speed at the electrode sheath edge was 0.47~cm/$\mu$s, approximately 10\% slower than the sound speed. 

The depleted EVDF has been modeled using a truncated Maxwellian 
\begin{equation}
f_e = \frac{\bar{n}_e}{\pi^{3/2} \bar{v}_{Te}^3} e^{-v^2/\bar{v}_{Te}^2} H(v_z + v_{\parallel,c}) 
\end{equation}
in \cite{baal:11} and \cite{loiz:11}. Here $v_{\parallel,c} = -\sqrt{2e(|\phi_b| + \phi)/m_e}$ is the speed associated with the cutoff velocity for a collision distribution at a potential $|\phi_b| + \phi$ from the boundary (the plasma is chosen for the reference potential). Also, note that in this model $\bar{n}_e$, and $\bar{v}_{Te}^2 = 2\bar{T}_e/m_e$ are parameters (not density or temperature -- see \cite{baal:11} for the moment expressions in terms of these parameters) and that $H$ is the Heaviside step function. Both references apply a fluid moment approach, and arrive at the generalized Bohm criterion
\begin{equation}
V_i \geq \bar{c}_s \biggl\lbrace 1 + \frac{\bar{v}_{Te}}{v_{\parallel,c}} \frac{\exp(-v_{\parallel,c}^2/\bar{v}_{Te}^2)}{\sqrt{\pi} [ 1 + \textrm{erf} (v_{\parallel,c}/\bar{v}_{Te})]} \biggr\rbrace^{-1/2}  ,
\end{equation}
where $\bar{c}_s \equiv \sqrt{\bar{T}_e/M_i}$. For this electron distribution, and assuming monoenergetic ions, the conventional kinetic Bohm criterion also provides the same equation \cite{baal:11}. 

These predict that modifications to Bohm's original criterion occur when $v_{\parallel,c}/\bar{v}_{Te} \lesssim 1$ (i.e., when the sheath potential drop is smaller than an electron temperature). For the wall sheath $v_{\parallel,c}/\bar{v}_{Te} \simeq \Delta \phi/T_e \simeq 20$, so any corrections from kinetic effects in the EVDF are expected to be negligible. However, for the wall sheath $v_{\parallel,c}/\bar{v}_{Te} \simeq 0.6$. This predicts an approximately 15\% reduction in the ion speed at electrode sheath edge of the electrode compared to the wall. This is also consistent with the PIC-MCC data within the accuracy of identifying the sheath edge. Future work will study smaller sheath potentials where such electron kinetic effects are predicted to be more significant. A complete absence of Deybe sheaths has been observed in recent particle simulations~\cite{camp:12}.


\section{Conclusions}

The Bohm criterion is remarkable for its ability to treat situations that differ widely from the assumed plasma model underlying the theory (monoenergetic ions and Boltzmann electron density). Kinetic generalizations seek to understand the broad applicability of this theory, and to understand its limitations, by treating arbitrary ion and electron velocity distributions. The conventional kinetic Bohm criterion is based on a $v_z^{-1}$ moment of the Vlasov equation. This provides some generalization, treating the case of Bohm's assumptions and a broader category of possible distribution functions. However, it is limited by the assumption of an infinitely thin sheath, which implies that no ions leave the sheath. This is quite restrictive since physical sheaths are all finite in extent, allowing some ions to escape. The $v_z^{-1}$ moment procedure makes this approach particularly sensitive to small deviations from the assumed collisionless Vlasov solutions at low energy.

An alternative formulation based on standard fluid moments of the kinetic equation (positive-exponent velocity moments) provides a robust generalization, but model distribution functions must be specified before the general formulation can be closed. Although most cases reduce to simply interpreting the fluid variables in Bohm's criterion in terms of the appropriate moments of the distribution functions, exceptions do exist. These demonstrate interesting scenarios where kinetic effects can modify Bohm's criterion, resulting in new physical effects such as subsonic ion flows at the sheath edge. 

The two examples considered highlighted the relative advantages of the fluid moment approach. The first considered ionization in the plasma and sheath. Since ions can leave the sheath in this situation when they are born with enough directed energy to do so, the conventional kinetic Bohm criterion cannot be applied.  However, the criterion calculated from the fluid moment approach predicts the fluid flow speed to be very close to the ion sound speed, supporting the robustness of Bohm's original criterion. The second example of probes biased near the plasma potential demonstrated a case where substantial depletion of the EVDF leads to modifications of Bohm's original criterion. The kinetic theory predicts that ions become subsonic as the ion sheath potential drop falls below the electron temperature. This prediction was supported by PIC-MCC simulations.

\section*{Acknowledgments}

The authors thank Prof.\ Scott Robertson for providing the code from \cite{robe:09}, which was used as a model to develop the numerical solutions in Sec.~\ref{sec:ionc}. This research was supported in part by the Office of Fusion Energy Science at the U.S.\ Department of Energy under contract DE-AC04-94SL85000, and in part by a University of Iowa Old Gold Summer Fellowship.



\end{document}